\documentstyle[aps,epsfig,floats]{revtex}
\newcommand {\be} {\begin{equation}} 
\newcommand {\ee} {\end{equation}} 
\newcommand {\Be}{\begin{eqnarray*}}
\newcommand {\Ee} {\end{eqnarray*}}
\newcommand {\bey} {\begin{eqnarray}} 
\newcommand {\eey} {\end{eqnarray}} 

\begin{document}

\begin{center}

\Large{\bf Energy transport in anharmonic lattices \\
close and far from equilibrium }\\

\vspace{0.5cm}

{\large Stefano Lepri}
\footnote{also Istituto Nazionale di Fisica della Materia, 
Unit\`a di Firenze} \\ 
{\small\it
Max-Planck-Institut f\"ur Physik Komplexer Systeme,
N\"othnitzer Str. 38, D-01187 Dresden, Germany\\
{\rm lepri@mpipks-dresden.mpg.de}\\
}

\vspace{0.3cm}
{\large Roberto Livi} $^{*\dagger}$\\
{\small\it
Dipartimento di Fisica dell'Universit\`a, 
L.go E.Fermi 2 I-50125 Firenze, Italy\\
{\rm livi@fi.infn.it}\\
}

\vspace{0.3cm}
{\large Antonio Politi} \footnote{
also Istituto Nazionale di Fisica Nucleare, Sezione di Firenze }\\
{\small\it
Istituto Nazionale di Ottica, 
L.go E.Fermi 6 I-50125 Firenze, Italy\\
{\rm politi@ino.it}\\
}

\end{center}

\date{\today}
\begin{abstract}
{The problem of stationary heat transport in the Fermi-Pasta-Ulam 
chain is numerically studied showing that the conductivity diverges in
the thermodynamic limit. Simulations were performed with time-reversible
thermostats, both for small and large temperature gradients. 
In the latter case, fluctuations of the heat current are shown to be 
in agreement with the recent conjectures of Gallavotti and Cohen
[Phys. Rev. Lett. {\bf 74}, 2694 (1995)].}

\vspace{0.2cm}
\vspace{0.4cm}

\noindent
{\it To be published in Physica D}
\end{abstract}

\section{Introduction}

The problem of heat conduction in one-dimensional insulating solids is very 
old. The celebrated Peierls' theory \cite{peierls} was successful in describing
the low-temperature dependence of the thermal conduction cofficient, clarifying
the basic role of Umklapp scattering induced by nonlinearity. However, 
for sufficiently high energies/temperatures, where the usual picture of weakly 
interacting phonons is no longer appropriate, one has to face a fully nonlinear
problem and a complete analytical solution seems hardly feasible.

Accordingly, in such a regime, the problem has been attacked several times 
by means of nonequilibrium molecular-dynamics (see Ref.~\cite{noi} and the 
bibliography therein). This amounts to solving numerically the microscopic 
(classical) equations of motion, with the principal goals of recovering the 
macroscopic heat-diffusion law and of measuring the corresponding transport 
coefficient. However, even this program turned out to be far from trivial, 
especially in one spatial dimension, where the interpretation of the numerical
analysis appears to be rather controversial.

The only clear results refer to some 1D toy models, 
characterized by a non-smooth interaction, like the so called ding-a-ling model 
\cite{dingling} and its simplified version studied in Ref.~\cite{robnick}. 
The remarkable success of such artificial systems in producing a 
``normal'' (i.e. length-independent) heat conductivity is claimed to follow 
from the strong inhibition of  coherent soliton propagation (see, however,
Ref.~\cite{Vis76} for a counterexample). Accordingly, the main 
mechanism at work is the diffusive transport of the energy induced by the 
underlying chaotic dynamics, which thus ensures the validity of the Fourier 
law.

As soon as one turns to the more realistic case of smooth interaction 
potentials, the scenario is definitely more ambiguouos. For instance, it is 
claimed that the conductivity of a diatomic Toda lattice is finite or 
diverging depending on the mass ratio \cite{JM89}~. For the Fermi-Pasta-Ulam 
(FPU) chain, the results reported in Ref.~\cite{noi} and in this 
paper do not show any evidence of a normal conductivity. In every case, 
there is no clear explanation of the observed phenomenology, even in the limit
of small applied gradients, so that it is worth reconsidering the whole problem.

This is not, however, our only motivation. Recently, the approach to 
non-equilibrium statistical mechanics through the introduction of 
time-reversible thermostats proved to be rather effective in several
test problems \cite{EM}. In fact, if a system, besides being time-reversible,
is ``sufficiently chaotic'', the tools developed for strictly 
hyperbolic systems allow for some statistical predictions even far 
from equilibrium \cite{GC95}. 
Relevant numerical tests have been performed on shear-flow
\cite{ECM93} and electrical conduction models \cite{BGG96}.
In the second part of the present work, we successfully test these
ideas also for the heat conduction problem.

More specifically, in section II we introduce the model equations, while
section III is devoted to a brief description of the microscopic quantities
of interest. Macroscopic properties of heat conduction are discussed in
section IV, and the predictions of the ``chaotic hypothesis'' are tested in
section V. The open problems and future perspectives are shortly summarized in
section V. Finally, the expression for the heat flux measured in the simulations
is derived in the Appendix.

\section{Models with time-reversible thermostats} 

We consider a chain of $N$ anharmonic oscillators, denoting with $q_l$ the 
displacement of the $l$-th particle from its equilibrium position. The
lattice has fixed boundary conditions, $q_0=q_{N+1}=0$~. In the bulk 
($l=2,\dots,N-1$), the equations of motion are purely hamiltonian, namely
\be
\label{f=ma}
\ddot q_l = f_l - f_{l+1} \quad ,
\ee
where $f_l=-V'(q_l-q_{l-1})$ is the force resulting from the interaction 
potential $V$ (the prime denotes derivative with respect to the argument)
acting between neighbouring particles. In the following we will consider the 
FPU potential
\be
V(x)\, =\, {x\over 2}^2+ \beta {x\over 4}^4
\ee
with $\beta=0.1$. We anticipate that the results described here in the
following do not appear to be peculiar of this choice of the potential.
A similar scenario has been found for different choices of $V$.

The first and the $N$-th oscillators interact with two heat reservoirs 
operating at different temperatures, $T_L$ and $T_R$, respectively (without 
loss of generality, we assume that $T_L>T_R$), in order to induce a heat flux 
through the chain. Their equations of motion are
\bey
\label{model}
&\ddot q_1 &= -\zeta_L\, \dot q_1 + f_1 - f_2 \nonumber \\
&\ddot q_N &= -\zeta_R\, \dot q_N + f_N - f_{N+1} \quad.
\eey 
The action of the thermostats is microscopically modelled by the ``thermal'' 
variables $\zeta_L$, $\zeta_R$, which evolve according to the Nos\'e-Hoover 
dynamical equations \cite{nose}
\bey
\label{nosehoover}
&&\dot \zeta_L \;=\; {1\over \Theta^2}\left({{\dot q_1}^2 \over T_L}
  -1\right) \nonumber\\
&&\dot \zeta_R \;=\; {1\over \Theta^2}\left({{\dot q_N}^2 \over T_R}
  -1\right) \quad ,
\eey
where the time $\Theta$ is the thermostat response time.
The above prescriptions imply that the kinetic energy of the boundary 
particles fluctuates around the imposed average value, thus simulating 
a ``canonical'' dynamics. In the limit case $\Theta \to 0$, one has the
so-called isokinetic (or Gaussian) thermostat: the kinetic energy is exactly 
conserved and the action of the thermal bath is properly described without the 
need to introduce a further dynamical variable, since $\zeta$ becomes an 
explicit function of the $\dot q$s \cite{EM}. We shall see that there is a price
to pay for the simplification of the dynamical equations: a larger
thermal resistance at the boundaries.

For all values of $\Theta$, the equations of motion are left invariant under 
time-reversal composed with the involution $\cal I$ defined as 
$\dot q_l\to -\dot q_l$, $\zeta\to -\zeta$ (notice that the $\zeta$s are 
momentum-like quantities \cite{nose})~. For a 
discussion of time reversibility in such thermostatted models and its 
implications to nonequilibrium statistical mechanics see Ref.~\cite{hoover}.

\section{Microscopic definition of the thermodynamic quantities}

Let us define the local energy density of the chain as
$h(x,t) = \sum_l h_l \delta (x-x_l)$, where $x_l = la + q_l$ is the position
of the $l$-th particle in a lattice of spacing $a$, while
\be
h_l := {p_l\over 2}^2+ {1\over 2}\left[ V(q_{l+1}-q_{l})+ V(q_l-q_{l-1})
\right]
\label{edens}
\ee
is the energy per particle (as usual, all the masses are set equal to 1
so that $p_l=\dot q_l$). If local equilibrium holds, the definition of
kinetic temperature stems from the local version of the ``virial theorem''
\be
\left\langle  p_l {\partial h_l\over \partial  p_l}\right\rangle=
\left\langle  q_l {\partial h_l\over \partial  q_l}\right\rangle= T_l
\quad ,
\label{viriale}
\ee
where $\langle\cdot\rangle$
denotes the time average over a sufficiently long trajectory. Assuming, as it 
is indeed the case of our model, that Eq. (\ref{viriale}) holds, we define 
$T_l=\langle p_l^2\rangle$. 

The local heat flux $j(x,t)=\sum_l j_l \delta (x-x_l)$ is implicitely 
defined by the continuity equation
\begin{equation}
\dot h(x,t) \,+\, {\rm div}\, j(x,t)\;=\; 0 \quad .
\label{continuity}
\end{equation}
The very definition of $j_l$ is non-trivial in strongly anharmonic systems 
i.e. at high energies/temperatures. The correct way to proceed
(see the Appendix and Ref.~\cite{Choq}) is to perform the spatial Fourier
transform of Eq.~(\ref{continuity}) and to expand the result in powers of
the wavenumber $k$. Neglecting all higher-order terms, as it is usually done
in hydrodynamics, one eventually obtains that the heat flux at the $l$-th
position is given by
\begin{equation}
j_l(t) \,=\,{1\over 2} \, a p_l\, \left(f_{l+1}+f_l\right) \quad ,
\label{flux}
\end{equation}
so that $p_l f_{l+1}$ has the simple interpretation of the flow of potential
energy from the $l$-th to the neighbouring particle. In the simulations, 
we have measured the time average $\langle j_l\rangle$ in the bulk. Notice
that the stationarity condition implies 
$\langle p_l f_{l+1} \rangle = \langle p_l f_l \rangle$, as it can be easily 
derived  from the condition $\langle\ddot q_l\rangle=0$.  

For later convenience we define the total flux through a chain or a subchain 
of $N$ particles, as the integral of $j(x,t)$, namely
\be
J(t) \;=\; {1\over N} \sum_l j_l(t) \quad.
\label{global}
\ee
At equilibrium, the instantaneous fluxes fluctuate ``randomly'' around zero.

\section{Macroscopic properties in the nonequilibrium state}

Here, we describe the outcome of our molecular-dynamics simulations,
performed with fixed values of the temperatures $T_R$ and $T_L$. With
this setting, larger numbers of particle correspond to smaller
temperature gradients (i.e. small external fields), so that, for 
$N \to \infty$, equilibrium dynamics is locally approached along the chain.
In other words, we are considering cases where the usual linear response or 
Green-Kubo theory is eventually applicable. Nevertheless, the problem 
remains highly nontrivial and deserves, as we will see, much attention.

The numerical simulations have been performed with several values of $N$ up
to 4096, monitoring both the kinetic temperatures and the heat fluxes.
In every case, after a suitably long transient, the system reaches a 
statistically stationary state, where each oscillator is in local equilibrium 
at a certain kinetic temperature and Eq.~(\ref{viriale}) is well verified. 

\subsection{The temperature field}

Whenever the imposed temperatures are different from one another, the 
microscopic dynamics
corresponds to a nonequilibrium macrostate characterized by a nonuniform 
temperature field along the chain (see Fig.~1). Long averaging 
reveals that the temperature gradient is rather smooth except at the 
boundaries, where thermal resistance effects may generate large
temperature jumps. Such effects turn out to depend on the response time
of the thermostats and are particularly relevant for small $\Theta$'s 
(see third panel in Fig.~1). One can understand this phenomenon by 
realizing that a fast reaction of the thermostats makes the dynamics of 
the end particles qualitatively different from that of the neighbouring 
ones. As one is interested in measuring the bulk contribution to the 
conductivity, it is important to minimize the
boundary effects, i.e. the temperature gaps observed at both extrema
\footnote{Similar problems arise also with random thermostats: 
ingeniuous tricks have been worked out to circumvent them \cite{robnick}.}.
This implies that $\Theta$ should be chosen as large as possible; however,
the larger is $\Theta$, the longer must last the simulations in order to
have reliable statistics. We find that the choice $\Theta = 1$ represents
a reasonable compromise.

As already observed in Ref.~\cite{noi}, the behaviour of the temperature 
profiles for different numbers of particles is well reproduced by the scaling 
Ansatz,
\be
T_l = T(l/N) \quad,
\ee
i.e., the shape of the profile is independent of $N$, if the chain length is 
rescaled to 1. 
This implies that the temperature field scales everywhere in the same
manner and one can choose equivalently any interval for measuring the
temperature gradient (provided that the interval is measured in fractions
of the total length).

The temperature profile in the bulk is not exactly linear (except for rather 
small $N$s), as one could expect from the stationary solution of the heat 
equation (namely the Fourier law). The nonlinearity of the profiles could 
at first be interpreted as an indication of a temperature-dependent 
conductivity, but this is actually not the case. In fact, simulations 
performed with as small temperature-differences as $T_L-T_R=4$ and for
$N=128$, still reveal clear deviations from linearity.

\begin{figure}
\hspace{.5cm}\centering\epsfig{figure=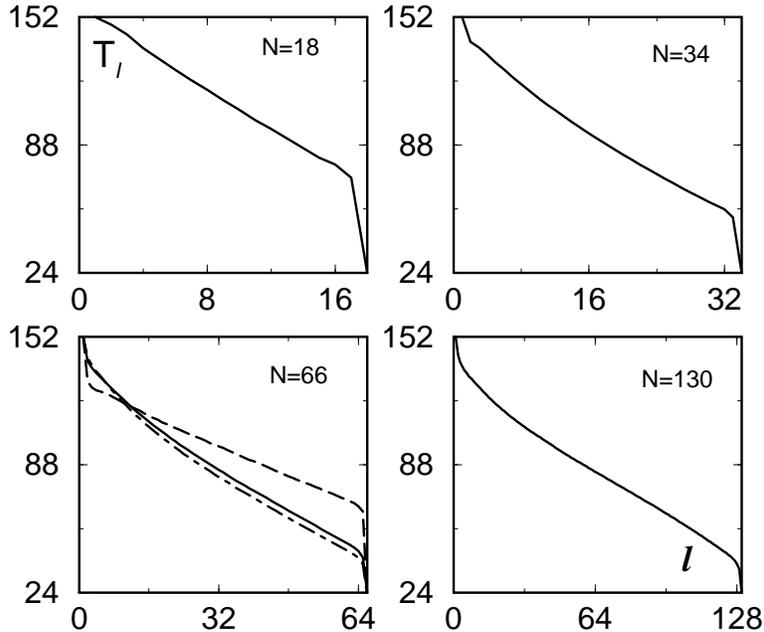,width=12cm}
\begin{minipage}[b]{15cm}
\caption{\baselineskip=12pt
Temperature profiles with $T_L=152$, $T_R=24$, $\Theta=1$, for 
different lattice lengths. Note that a nonlinear profile sets in for the
larger $N$ values. For $N=66$ we report the temperature fields for 
two other values of the response time, namely $\Theta=0.1$ (dashed line)
and $\Theta=10$ (dot-dashed line).}
\end{minipage}
\end{figure}
\subsection{Thermal conductivity}

Another relevant macroscopic feature is the onset of a constant, on average,
heat flux, namely $\langle j_l\rangle=j$, for every $l=2,\dots,N-1$
(the flux through the oscillators in contact with the heat reservoirs 
must account also for the energy exchange with each thermostat - see the 
following subsection).

The thermal conductivity is defined (to the lowest order in the applied 
gradient) as the ratio 
\begin{equation}
\kappa = {j \over dT/dx} \quad .
\label{conduc}
\end{equation}
Our main result is that $j$, which vanishes as $N\to\infty$, approaches 0 
more slowly than the temperature gradient, thus implying a diverging
conductivity in the thermodynamic limit. In fact, the simulations do 
reveal that the heat flux scales as $j\propto N^{-\alpha}$, with $\alpha$ 
definitely smaller than 1. Accordingly (see Fig.~2), $\kappa$ diverges as 
$N^{1-\alpha}$, since the scaling behaviour of the profile implies that 
$dT/dx$ is proportional to $(T_L-T_R)/N$. A careful scrutiny of the data in
Fig.~2 reveals a sort of crossover from $\alpha \approx 0.55$, for $N<250$
to $\alpha \approx 0.62$ for larger numbers of particles. 

It is worth mentioning that a similar behaviour of the conductivity is found 
also by using other thermostatting schemes, such as stochastic interactions
with a gas of Maxwellian particles. For comparison, in the inset of Fig.~2,
we report the data taken from \cite{giapponesi}. Although they refer to 
different temperatures ($T_L=1500$, $T_R=150$, if expressed with reference 
to the same value of $\beta$ here employed), the scaling behaviour is 
approximately the same as ours.

Therefore, we are forced to conclude that, at least for the considered
system sizes, Fourier law is not satisfied and chaoticity is not 
sufficient to ensure its validity. 

\begin{figure}
\hspace{.2cm}\centering\epsfig{figure=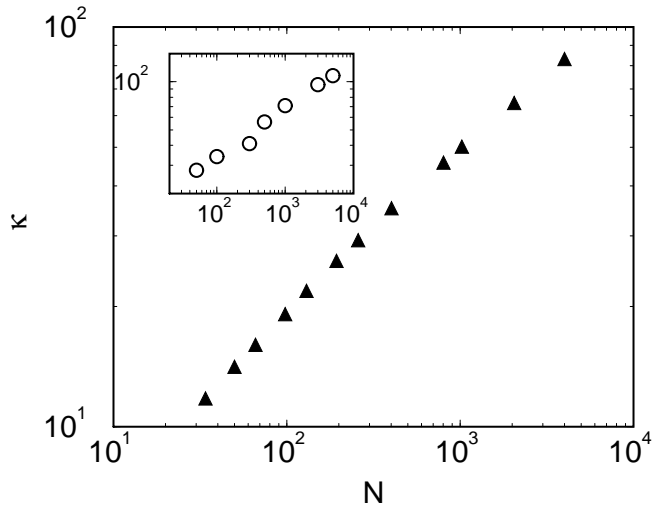,width=10cm}
\begin{minipage}[b]{15cm}
\caption{\baselineskip=12pt
Behaviour of the thermal conductivity, Eq.~(\protect\ref{conduc}), 
as a function of the lattice length $N$ for $T_L=152$, $T_R=24$. The flux $j$ 
is computed averaging over one long trajectory ($\approx 10^6$), 
started from random initial conditions and after discarding a transient 
($\approx 10^4$). The inset shows the results of Ref.~\protect\cite{giapponesi}. }
\end{minipage}
\end{figure}

\subsection{Entropy production}

The energy flux at the chain ends is simply given by
\be
j_{L,R} =  -\langle \zeta_{L,R} \, \dot q^2_{1,N} \rangle
= - \langle\zeta_{L,R}\rangle T_{L,R} \quad,
\label{fluxbath}
\ee
where the second equality is obtained from the condition
$\langle d\zeta^2_{L,R}/dt\rangle=0$. In the stationary regime, the balance
between the ingoing and outgoing fluxes implies that $j_L = -j_R$ which,
in turn, implies that $\langle \zeta_L \rangle$ and
$\langle \zeta_R \rangle$ must have opposite signs. This rather obvious
relationship, stemming from energy conservation, has a somehow surprising
meaning, if interpreted from the point of view of dynamical equations. In
fact, a negative $\langle \zeta_L \rangle$ (the flux must be obviously
positive in the left-end, which is in contact with the hotter reservoir)
means an expansion of volumes rather than a dissipation! The apparent
anomaly is immediately clarified by noticing that the system in the whole is
globally dissipative as $\gamma = \langle \zeta_R+\zeta_L\rangle$ turns out
to be positive in all simulations that we have performed.
This is also consistent with a theorem recently proven by Ruelle \cite{R96}
for time-reversible systems and expresses the intuitive notion that
if the energy is conserved on the average, it is not possible that volumes
steadily diverge. What is not a priori trivial is that a finite dissipation
spontaneously arises as soon as a non-equilibrium stationary state sets in
\cite{GC95,ECM93}. 

Eq.~(\ref{fluxbath}) is also susceptible of an interpretation in terms of
entropy production. By subtracting from one another the two expressions of
the fluxes and noticing that $j = j_L = -j_R > 0$, one obtains
\begin{equation}
\langle \zeta_L\rangle + \langle \zeta_R\rangle \,=\, 
j \left( {1\over{T_R}} \, - \, {1\over{T_L}}  \right) \quad.
\label{balance}
\end{equation}
which can be interpreted as a balance relation for the global entropy
production. In fact, according to the principles of irreversible
thermodynamics, the local rate of entropy production $\sigma$ in the bulk 
is given by
\begin{equation}
\sigma(x)\,=\, j {d\over dx}\left( {1\over T(x)} \right) \quad .
\label{sigma}
\end{equation} 
Upon integrating Eq.~(\ref{sigma}), the r.h.s of Eq.~(\ref{balance}) is
obtained, which can thus be interpreted as the global production rate of
entropy in the bulk. On the other hand, according to general arguments on
reversible thermostats \cite{GC95}, the l.h.s. of Eq.~(\ref{balance}) can be
identified with the entropy production from the heat baths.
Eq.~(\ref{balance}) has been accurately verified in a wide range of
temperatures.

\section{Chaotic Hypotesis and large fluctuations}

We now turn our attention to far-from-equilibrium regimes, where the
applied gradient is very large, i.e. to the case of short chains with
large temperature differences at the boundaries.
Our aim is to check the validity of the fluctuation theorem, recently
proposed by Gallavotti and Cohen \cite{GC95}, and carefully tested in the 
periodic Lorentz gas with an electric applied field \cite{BGG96}.
The test is rather crucial for at least two reasons: (i) at variance with
most of the cases considered in literature, here only the boundary particles
are thermostatted \cite{galjsp}; (ii) it is not a priori clear that the
{\it chaotic hypothesis}, one of the key assumptions for the validity of
the theorem, applies to the present case. In fact, it is far from obvious
that a dynamical system such as an FPU chain, displaying a slow convergence 
to equipartition at low temperatures, is ``Anosov-like'' in the thermodynamic
limit.

The fluctuation theorem essentially connects the probability of positive to
that of negative values of the entropy production. Although we recommend the 
reader to consult \cite{GC95} for a detailed exposition of the theorem, here 
we summarize the main lines of the proof to give a flavour of the various 
steps. The starting observation is that any two regions in phase 
space\footnote{Technically speaking, we should refer to a sufficiently small 
element of a Markov partition.}, mutually related by the involution $\cal I$, 
are characterized by opposite entropy values $\sigma$ and $-\sigma$ (the 
entropy changes sign upon applying the involution $\cal I$). The second key 
point is that the existence of a Sinai-Ruelle-Bowen measure implies that the 
probability of one such region $R$ is proportional to the product of the 
expanding multipliers (over a suitable time lag). As a consequence, invariance 
under time-reversal ensures that the probability to observe $- \sigma$ in $R$ 
is also proportional to the product of the inverse of the contracting 
multipliers in ${\cal I}(R)$. This fact implies that the ratio of the 
probability of observing $\sigma$ to that of $-\sigma$ can be reduced to the 
product of expanding and contracting multipliers, all taken at the point 
where $\sigma$ is actually observed. This product is, in turn, nothing but the 
volume contraction or, equivalently, the exponential of the entropy production.

From Eq.~(\ref{balance}), one can see that, apart from an irrelevant
multiplicative constant (the external-field amplitude), the heat flux $j$ is 
equivalent to the entropy production. In order to make the analogy with
the previous study even more stringent, we have, however, preferred to
consider the global heat flux as defined in Eq.~(\ref{global}) as the
latter quantity corresponds to a spatial average over all oscillators
(let us indeed recall that in many previous studies the entropy production
is an extensive quantity, proportional to the number of degrees of freedom). 
This change of definition does not affect the symmetry properties of $J$, so 
that the fluctuation theorem applies (or not) independently of this 
modification.

Therefore, our starting point is the finite-time average of the global heat 
flux (proportional to the entropy production), 
\be
\langle J \rangle_\tau \;=\; {1\over \tau}\int_t^{t+\tau} J(t')\, dt'
\quad ,
\ee
where the average is performed over a sufficiently long trajectory to 
ensure a good accuracy of the underlying Markov approximation (see again
Ref.~\cite{GC95}). We then compute the probability distribution $P_\tau$ of
the variable
\be
z \;=\; {{\langle J \rangle_\tau }\over {\langle J \rangle_\infty}}
\ee
where $\langle J \rangle_\infty$, denoting the ``infinite time'' average 
of the global flux, coincides with the stationary average flux $j$ introduced 
in the previous sections. The Gallavotti-Cohen conjecture then reads as
\be
\label{flthm}
\ln{P_\tau(z)\over P_\tau(-z)} \;=\; 
\tau z j\left({1\over T_R}-{1\over T_L}\right)
\quad .
\ee
We have performed simulations for $N=14$, $T_L=120$, $T_R=56$, $\Theta=1$
and several values of $\tau$ ranging from 5 to 80. At variance with the
results of Ref.~\cite{BGG96}, the bell-shaped distribution $P(z)$ (Fig.~3)
is clearly not Gaussian: both of its tails approach zero exponentially for
large values of $|z|$, but with different rates. Upon increasing $\tau$, the
central part of $P_\tau (z)$ approaches more and more a Gaussian
shape as one would expect from the increasing statistical independence of
the data. On the other hand, reliable numerical measures become more and
more difficult: for instance, already for $\tau=80$, negative values of $z$
are much too rare to permit the evaluation of the distribution over a
reasonable integration time. Nevertheless, the linear behaviour predicted by
Eq.~(\ref{flthm}) is indeed observed, as shown in Fig.~4, where the
numerical points are superimposed to the theoretical prediction.

\begin{figure}
\hspace{.5cm}\centering\epsfig{figure=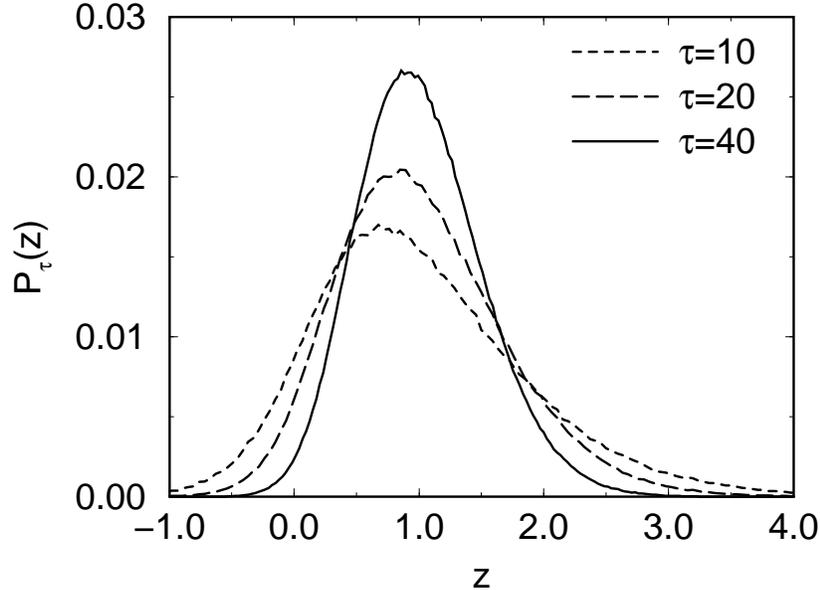,width=12cm}
\begin{minipage}[b]{15cm}
\caption{\baselineskip=12pt
The distribution $P_\tau(z)$ for different values $\tau$. The
chain lenght is $N=14$ and the boundary temperatures are  
$T_L=120$, $T_R=56$ and $\tau_0=50$. The value of the average flux 
is $j=30.67\pm0.01$.}
\end{minipage}
\end{figure}
\begin{figure}
\hspace{.5cm}\centering\epsfig{figure=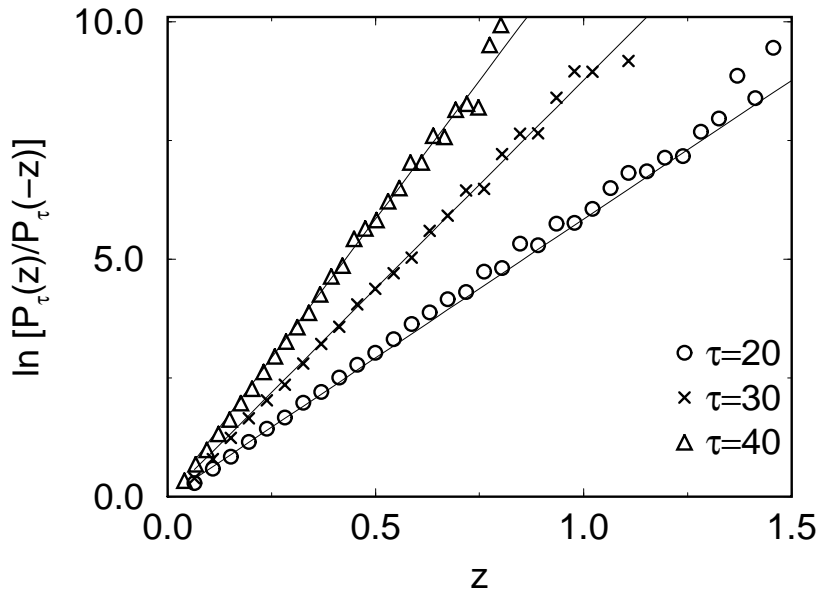,width=12cm}
\begin{minipage}[b]{15cm}
\caption{\baselineskip=12pt
Test of the fluctuation theorem, for the same parameters as
in Fig. 3 and different values of the time $\tau$. The solid lines are 
the predictions of Eq.~(\protect\ref{flthm}).}
\end{minipage}
\end{figure}

\section{Conclusions and Perspectives}

In this paper we have shown that the fluctuation theorem recently proposed 
by Gallavotti and Cohen \cite{GC95} is successfully verified for a chain of 
FPU oscillators in contact with time-reversible thermostats at the boundaries. 
It is worth stressing that this is a very different context with respect to 
the ones previously considered in the literature \cite{ECM93,BGG96}. In this
sense, our results definitely enforce the generality of this approach in 
describing the thermodynamics of nonequilibrium stationary states in terms of 
the Sinai-Bowen-Ruelle probability distribution.

Furthermore, we have found that heat conductivity appears to diverge in the
thermodynamic limit. This is at variance with results obtained in
\cite{dingling,robnick} with reference to the so-called ding-a-ling model.
The origin of the divergence of $\kappa$ in FPU seems to be traceable to
the existence of quasi-conserved long-wavelength modes \cite{casar} 
supporting almost ballistic transport along the chain. Such a feature is
indeed generic for Hamiltonian models with smooth nearest-neighbour 
coupling like (\ref{f=ma}), but it is clearly absent in models such as 
ding-a-ling which cannot sustain long-wavelength modes.

We plan to proceed towards a more detailed comprehension of heat conduction 
by exploring the following routes:

\begin{itemize}
\item Dependence of transport properties on the physical settings;
for example, by changing the number of particles in thermal contact
with the reservoirs.
 
\item  Measurement of the thermal conductivity in the framework of
Green-Kubo linear response theory with equilibrium simulations.

\item Application of the same approaches to different models of nonlinear
chains of oscillators, including the extension to two and three dimensions.

\end{itemize}

In particular, we want to stress the importance of the last point since it 
is not unlikely that macroscpic validity of the Fourier law is generally 
ensured only in some higher dimension. After all, statistical mechanics is 
full of phenomena that do depend on the dimensionality.

\acknowledgements
We thank G.~Gallavotti for suggesting the present study and E.G.D.~Cohen 
for useful discussions. We are also indebted with Ph.~Choquard for having 
clarified to us the definition of the heat flux.
\appendix
\section{}

In this Appendix we detemine the expression of the heat flux
for a 1d chain, where the neighbouring particles interact via a
generic nonlinear potential, such that the resulting equations of 
motion are of the form (\ref{f=ma}). For simplicity we assume 
that the chain is infinite.

Let us write the energy density and the flux as Fourier integrals
\be
\tilde h(k,t) = \int h(x,t) e^{-ikx} dk \quad \quad
\tilde j(k,t) = \int j(x,t) e^{-ikx} dk \quad \quad,
\ee
so that Eq.~(\ref{continuity}) becomes
\be
{d \tilde h\over dt}+ik\tilde j = 0 \quad.
\ee
We now rewrite the last equation, by splitting the heat flux into two
different contributions
\be
\tilde j \;=\; \tilde j^{(1)}+\tilde j^{(2)} \quad,
\ee
such that, by the explicitely computing the time-derivative of $\tilde H$,
it is recognized that
\be 
\tilde j^{(1)} = \sum_l\, p_l h_l \, e^{-ikx_l} \quad ,
\ee
while
\be 
-ik\tilde j^{(2)} = \sum_l\, \dot h_l e^{-ikx_l} = 
{1\over2} \sum_l \, f_l (p_l+p_{l-1})\,(e^{-ikx_l}-e^{-ikx_{l-1}})
\quad .
\ee
As we want to determine $\tilde j^{(2)}$ to the lowest order in $ik$, we
expand the exponentials as
\be 
e^{-ikx_l}-e^{-ikx_{l-1}} 
= e^{-ikx_l} \,\left(-ik(x_l-x_{l-1})+{\cal O} (k^2)\right)
\quad ,
\ee
end then
\be 
\tilde j^{(2)} \approx 
{1\over2} \sum_l \, f_l (p_l+p_{l-1})\,(x_l-x_{l-1}) e^{-ikx_l}\quad .
\ee
Summing the two terms and recalling the definition of $j$, 
it can be realized that (after some further index manipulation)
\be
j_l \;=\;  p_l h_l + {1\over2}(q_l-q_{l-1})(p_l+p_{l-1})f_l +
{a\over2} p_l (f_l + f_{l+1}) \quad .
\ee
With similar calculations, it can be shown that the first two terms
in the r.h.s. correspond to order-$k^2$ terms in the Fourier
expansion so that they can be neglected and Eq.~(\ref{flux}) is finally
obtained.


\newpage
\begin{thebibliography}{99}

\bibitem{peierls} R.E. Peierls, {\it Quantum theory of solids}, Oxford
University Press, London (1955).
\bibitem{noi} S. Lepri, R. Livi, A. Politi, Phys. Rev. Lett.
{\bf 78} (1997) 1896.
\bibitem{dingling} G.~Casati, J.~Ford, F.~Vivaldi, W.M.~Visscher,
Phys. Rev. Lett. {\bf 52} (1984) 1861 .
\bibitem{robnick} T.Prosen, M. Robnik, J. Phys. A {\bf 25} (1992) 3449. 
\bibitem{Vis76} W.M.~Visscher, in {\it Methods in Computational Physics}
{\bf 15}, Academic Press, New York (1976) 371.
\bibitem{JM89} E.A.~Jackson, A.D.~Mistriotis, J. Phys. Condens. Matter
{\bf 1} (1989) 1223 . 
\bibitem{EM} D.J.~Evans, G.P.~Morriss, {\it Statistical Mechanics of
Nonequilibrium Liquids}, Academic Press, San Diego (1990).
\bibitem{GC95} G. Gallavotti, E.G.D.~Cohen, J. Stat. Phys. {\bf 80} 
(1995) 931 and Phys. Rev. Lett. {\bf 74} (1995) 2694.
\bibitem{ECM93} D.J.~Evans, E.G.D.~Cohen, G.P.~Morriss, Phys. Rev. Lett.
{\bf 71} (1993) 2401.
\bibitem{BGG96} F. Bonetto, G.Gallavotti, P.Garrido, Physica D {\bf 105}
(1997) 226.
\bibitem{nose} S.~Nos\'e, J. Chem. Phys. {\bf 81} (1984) 511;
W.G.~Hoover Phys. Rev. A {\bf 31} (1985) 1695. 
\bibitem{hoover}B.L.~Holian, W.G.~Hoover, H.A.~Posch, Phys. Rev. Lett.
{\bf 59} (1987) 10.
\bibitem{Choq} Ph.~Choquard, Helvetica Physica Acta, {\bf 36} (1963) 415.
\bibitem{giapponesi} H.~Kaburaki, M.~Machida, Phys. Lett. A {\bf 181} 
(1993) 85. 
\bibitem{R96} D.~Ruelle, J. Stat. Phys. {\bf 85} (1996) 1.
\bibitem{galjsp} See, e.g., G.~Gallavotti, J. Stat. Phys. {\bf 84} (1996) 899 
for another such example.
\bibitem{casar} C.~Alabiso, M.~Casartelli, and P.~Marenzoni, J. Stat. Phys.
{\bf 79} (1995) 451. 

\end{thebibliography}
\end{document}